\documentstyle[11pt,newpasp,epsf,twoside]{article}
\index{pulsars!emission models}
\index{pulsars!polar caps}

\begin{document}
\title{A Resonant-Mode Model of Pulsar Radio Emission}
 \author{M. D. T. Young}
\affil{School of Physics, University of Western Australia, 35 Stirling 
Highway, Crawley WA 6009, Australia}

\begin{abstract}
It is argued that the polar gap and flux tube in the pulsar
magnetosphere act as a resonant cavity/waveguide system which is
excited by oscillations in the primary beam current and accelerating
potential.  The modes will be converted, probably scattered, to
produce radio beams in the frequency range of those observed.  
\end{abstract}

\section{Overview of the Model}

We live in a resonant cavity bounded below by the earth and above by
the ionosphere.  Lightning strikes excite low-frequency
electromagnetic ``Schumann resonances'' in this cavity, as predicted
by Schumann (1952), with a fundamental frequency of about 8~Hz.  Such
behaviour should also occur above the magnetic poles of pulsars due to
the well-defined boundaries found there.

One of these boundaries is the stellar surface which, because of the
star's large electron number density, has a conductivity perpendicular
to the magnetic field comparable to that of copper; the boundary
condition there can be taken to be \(E_{\perp}=0\), where
perpendicular, $\perp$, and parallel, $\|$, are taken relative to the
stellar magnetic field.  Second, there is the boundary between the
open and closed magnetic field lines, which forms a diverging tube. 
On the closed field lines there is a non-neutral plasma at the
Goldreich-Julian charge density (Goldreich \& Julian; 1969), which is
co-rotating with the star.  The strong stellar magnetic field,
\(B_{\ast}\sim10^{12}\)~G, restricts the particle motion to be
essentially along the field lines, but this still allows them to
respond to an applied \(E_{\|}\).  The plasma can maintain
\(E_{\|}=0\) on the tubular boundary at frequencies up to the plasma
frequency, which is
\(\sim{}(7.5\mbox{~GHz})\,(B_{\ast}/10^{12}\mbox{~G})^{1/2}(0.1\mbox{~s}/P)^{1/2}\)
near the stellar surface.  On the open field lines particles are
accelerated away from the star by a non-zero \(E_{\|}\) to
ultra-relativistic speeds, creating $e^{-}/e^{+}$ pairs at a
pair-formation front, a third boundary.  Beneath the front is the
low-density acceleration zone, or polar gap; above it the pairs stream
up the tube away from the star forming a dense relativistic secondary
plasma.  The consequences of a second front forming the gap's base
(e.g. Harding \& Muslimov, 1998) are also being considered, but will
not be discussed here.

Sturrock (1971) noted that the current of particles up the tube might
oscillate due to intermittent screening of the accelerating field at
the surface of the star by positrons returning from the pair-formation
region, and this idea remains plausible.  Assume for now that the
current has an oscillating component; either this oscillation is in
the form of the ``sparking'' of Ruderman \& Sutherland (1975), or it
occurs as an AC component of the DC space-charge limited flow from the
surface.  This assumption will be revisited below.

\begin{figure}
   \centerline{\epsfxsize=\textwidth,\epsfbox{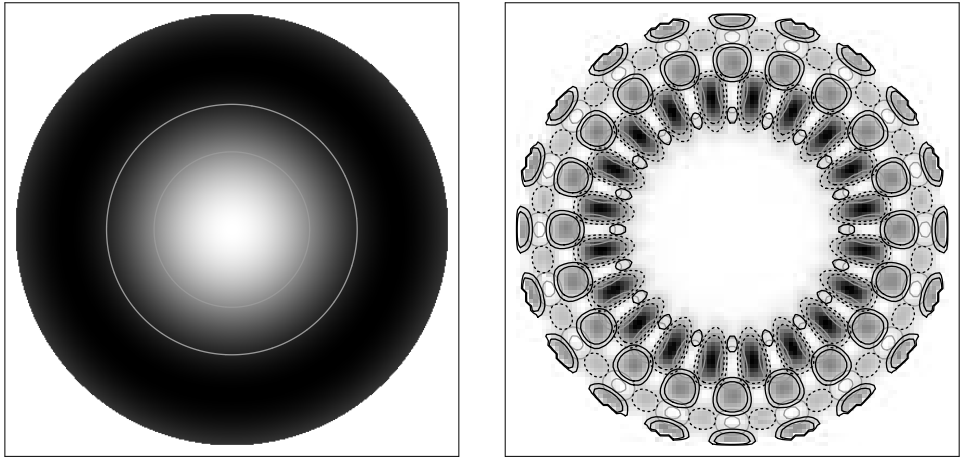}}
   \caption{Plots of the energy densities, $U_{\perp}$ and $U_{\|}$,
   of two modes: \emph{left} \((m,n)=(0,1)\), \emph{right}
   \((m,n)=(10,2)\).  The shading shows $U_{\perp}$, with maxima in
   black.  The gray contours delineate regions at the center of which
   $U_{\|}$ is maximum; the black and dashed contours delineate where
   $E_{\perp}$ has one of two orthogonal polarizations.}
   \label{fig:1}
\end{figure}

The AC component of current/potential will generate large-amplitude
electromagnetic waves~(Qiao, 1988) which will be essentially
transverse magnetic (\(B_{\|}\approx0\)) and have frequencies set by
the light travel time across the gap height, \(h\sim10^{2\pm1}\)~m, to
be around 0.1--10~MHz.  Taking account of the boundary conditions
discussed above, the waves will propagate in the tube as waveguide
modes.  Ignoring stellar rotation and vacuum polarization, and
assuming an infinite stellar magnetic field---so that the dielectric
tensor of the plasma is diagonal---the tube near the star supports
electromagnetic waveguide modes which in flat spacetime are generally
well approximated by the transverse magnetic modes of a cylindrical
waveguide of arbitrary cross-section.  In a circular cylinder the
modes are specified by two mode numbers $m$ and $n$, and examples are
shown in Figure~\ref{fig:1}.  Taking account of the change in the
dielectric tensor at the pair-formation front, the modes are equally
applicable both below and above the front.

The waveguide mode impedances change at the pair-formation front so
modes in the gap will be partially reflected from this boundary, and
since the modes will also be reflected from the stellar surface, the
gap can act as a resonant cavity.  I have studied the dispersion
relation of the plasmas below and above the front for various
configurations and will present my findings elsewhere.  Resonances are
possible in the gap at frequencies just above the cutoff frequency of
a vacuum filled cavity, though the dispersion relation differs from
that of the vacuum case.  In the secondary plasma above the front
there are two or three modes that can carry power up the tube; all
have different wavelengths, but the same low frequency as the
resonance in the gap.  Preliminary calculations indicate that the
oscillating $E_{\|}$ in the wind just above the gap may be many times
the oscillating $E_{\|}$ in the gap so that, even if the AC component
of field in the gap is a small fraction of the DC component, the
oscillating electric fields above the gap could be large.

When the driving frequency equals the frequency of a particular mode,
and the source oscillations are positioned where that mode has
antinodes, then that mode will be excited in the gap.  If the source
locations were to be confined symmetrically to any particular areas of
the polar cap, then we might suppose these to be either near the tube
axis, or perhaps close to the wall where a return current may flow. 
The former would tend to excite the \(m=0\) modes, and the latter,
high-order $m$ modes.

There are at least two phenomena that may provide a feedback mechanism
in the gap.  The first is the resonant field itself; energy pumped
into the field by the current in turn acts back on the current twice
per cycle.  Secondly the large $E_{\|}$ above the front may accelerate
particles of charge opposite to the primary beam current back into the
gap; this charge would then cross the gap in time $h/c$ to provide
increased screening at the base of the gap.  These phenomena would
tend to amplify the current/potential oscillations making the gap
unstable to resonance, and supporting the assumption made above of an
AC component.  Also the modes will be coupled via the current, and
this raises the possibilities of mode-locking, intermittency and
chaos, all of which will be discussed elsewhere.

The plasmas in the tube will posses a spatio-temporal sub-structure,
in particular above the antinodes of the resonant modes.  Perhaps this
structure together with the oscillating fields in the tube go on to
produce the observed radio emission via some sort of instability
mechanism (see Melrose, this volume, for a review).  However,
preliminary calculations indicate that at least one of the modes in
the tube will be scattered to the observed radio frequencies by the
inhomogeneities in the wind.  With increasing altitude in the tube,
the plasma density decreases and the tube diameter increases altering
the local dispersion relation, and this determines the range of
altitudes in which the scattering can occur.

Several features of pulsar radio emission have a natural
interpretation in the model.  The stability of mean pulse profiles
will be explained if the same modes are continuously excited, or there
is rapid switching between a fixed subset of modes.  If switching
between modes occurs on timescales of several pulses or more, it will
be recognized as the observed profile changes.  Nulling will be
observed if the mode has a low amplitude in the line of sight, or if
the resonance (and hence the emission) temporarily ceases---perhaps
because the pair-formation fronts become very irregular or the driving
oscillations cease as the primary beam current becomes steady.  The
model can produce short timescale intensity variations due to a few
factors, including the inhomogeneity in the plasmas and the finite
$Q$-factor of the cavity, and may explain the origins of
microstructure.  The \(m=0\) modes may correspond to the emission
usually denoted as `core' (Rankin, 1983; Lyne \& Manchester, 1988),
with the higher order $m$ modes yielding `conal' emission; indeed, for
\(m\approx{}7\mbox{--}10\) various modes or combinations of modes have
concentric rings of antinodes with a ratio of angular widths very
close to those determined for the inner and outer cones of Rankin
(1993).

The mode plots in Figure~\ref{fig:1} can also be directly interpreted
as polar emission maps, and show similarities to those of Deshpande \&
Rankin (1999) and others.  Furthermore, the polarization of the
scattered emission is determined by the polarization of the underlying
mode; since antinodes adjacent in the tube cross-section have
orthogonal polarization this gives a direct mechanism for the observed
orthogonal modes, whilst scattering of the $E_{\|}$ component would
tend to produce unpolarized emission, perhaps with some polarization
in the plane of curvature of the magnetic field lines.

Provided that the modes are not pinned to structure in the tube's
wall, the modes will be free to rotate about its axis and this
will be observed as subpulse drift.  For instance, this should be the
case when the inclination angle $\chi$ between a pulsar's rotation and
magnetic axes is small, making the tube's cross-section approximately
circular; in turn this might explain why pulsars exhibiting
quasi-stable subpulse drift generally have small $\chi$.

If it is also true that the modes are not pinned to the AC current
beams, but rather that the current beams follow the mode antinodes,
then the modes will rotate due to stellar rotation and general
relativistic frame dragging.  A standing wave can be viewed as a
distribution of oscillators in space, all of which are synchronized,
and to achieve this synchronization in some set of frames requires
that \(c^{-1} \oint (g_{0\alpha}/g_{00}) dx^{\alpha}=0\).  Then it is
found that the modes, whatever their form, must rotate with a
frequency
\(\left(1-{2}{G}I/{c^2}{r^{3}}\right)\cos\left(\chi\right)P^{-1}\)
where $r$ is the distance from the center of the star, and $I$ is the
stellar moment of inertia.  The sense of this rotation relative to the
stellar surface is opposite to that of the stellar rotation at the
rotational pole closest to the gap.  This does not reproduce the long
circulation times found by Deshpande \& Rankin (1999) and others and
this failure will be discussed elsewhere.  If frame-dragging and
stellar rotation are not the cause of subpulse drift, then it suggests
that the rotating modes are pinned to current density columns which
circulate about the axis, probably due to
\(\mathbf{E\times{}B}\)-drift.

\acknowledgments

I thank R. Burman and R. Manchester for many useful suggestions, and
UWA, the ATNF and my wife for financial support.

\end{document}